\documentclass[aps,prb,twocolumn,superscriptaddress,showpacs]{revtex4}

\usepackage{amsmath}
\usepackage[dvips]{graphicx}
\usepackage{color} %added by CFB

\newcommand{\vv}{{\bf {v}}}
\newcommand{\rr}{{\bf {r}}}
\newcommand{\pp}{{\bf {p}}}

\newcommand{\sss}{{\bf {s}}}

\newcommand{\ep}{{\mbox{$\epsilon_p$}}}
\newcommand{\eF}{{\mbox{$\epsilon_F$}}}

\newcommand{\pF}{{\mbox{$p_F$}}}

\newcommand{\vg}{{\mbox{$v_g$}}}
\newcommand{\bnabla}{{\mbox{\boldmath $\nabla$}}}

\begin{document}

\title{Visualization of quantum turbulence in superfluid $^3$He-B: Combined numerical/experimental study of Andreev reflection}

%USE WITH REVTEX4

\author{V.~Tsepelin}
\email[Electronic Address: ]{v.tsepelin@lancaster.ac.uk}
\affiliation{Department of Physics, Lancaster University, Lancaster, LA1 4YB, United Kingdom,}

\author{A.\,W.~Baggaley}
\affiliation{Joint Quantum Centre Durham-Newcastle, and School of Mathematics and Statistics, Newcastle University, Newcastle upon Tyne, NE1 7RU, United Kingdom,}

\author{Y.\,A.~Sergeev}
\email[Electronic Address: ]{yuri.sergeev@ncl.ac.uk}
\affiliation{Joint Quantum Centre Durham-Newcastle, and School of Mechanical and Systems Engineering, Newcastle University, Newcastle upon Tyne, NE1 7RU, United Kingdom,}

\author{C.\,F.~Barenghi}
\affiliation{Joint Quantum Centre Durham-Newcastle, and School of Mathematics and Statistics, Newcastle University, Newcastle upon Tyne, NE1 7RU, United Kingdom,}

\author{S.\,N.~Fisher}
\thanks{Deceased}
\affiliation{Department of Physics, Lancaster University, Lancaster, LA1 4YB, United Kingdom,}

\author{G.\,R.~Pickett}
\affiliation{Department of Physics, Lancaster University, Lancaster, LA1 4YB, United Kingdom,}

\author{M.\,J.~Jackson}
\affiliation{Faculty of Mathematics and Physics, Charles University, Karlovu 3, 121 16, Prague 2, Czech Republic,}

\author{N.~Suramlishvili}
\affiliation{Department of Mathematics, University of Bristol, Bristol, BS8 1TW, United Kingdom.}

\date {\today}

%%%%%%%%%%%%%%%%%%%%%%%%%%%%%%%%%%%%%%%%%%%%%%%%%%%%%%%%%%%%%%%%%%%%%%%%%%%%%%%%%%%%%%%%%%%%

\begin{abstract}
We present a combined numerical and experimental study of Andreev scattering from quantum turbulence in superfluid $^3$He-B at ultralow temperatures. We simulate the evolution of moderately dense, three-dimensional, quasiclassical vortex tangles and the Andreev reflection of thermal quasiparticle excitations by these tangles. This numerical simulation enables us to generate the two-dimensional map of local Andreev reflections for excitations incident on one of the faces of a cubic computational domain, and to calculate the total coefficient of Andreev reflection as a function of the vortex line density. Our numerical simulation is then compared with the experimental measurements probing quantum turbulence generated by a vibrating grid. We also address the question of whether the quasiclassical and ultraquantum regimes of quantum turbulence can be distinguished by their respective total Andreev reflectivities. We discuss the screening mechanisms which may strongly affect the total Andreev reflectivity of dense vortex tangles. Finally, we present  combined numerical-experimental results for fluctuations of the Andreev reflection from a quasiclassical turbulent tangle and demonstrate that the spectral properties of the Andreev reflection reveal the nature and properties of quantum turbulence.
\end{abstract}

\maketitle

\section{Introduction}
\label{sec:intro}

A pure superfluid in the zero temperature limit has no viscosity. The superfluid is formed by atoms which have condensed into the ground state. The condensate atoms behave collectively and are described by a coherent macroscopic wavefunction. In liquid $^3$He and dilute Fermi gases, the condensate is formed by pairs of atoms known as Cooper pairs. For most superfluids, including $^3$He-B, the superfluid velocity \vv\,  is proportional to the gradient of the phase of the wavefunction, $\bnabla\theta$. Thus the superfluid flow is irrotational $\bnabla\times\vv ={\bf 0}$ (more complex superfluids such as $^3$He-A are not considered here). Superfluids can however support line defects on which $\bnabla\times\vv$ is singular. These defects are quantum vortices. The phase $\theta$ of the wavefunction changes by $2\pi$ around the vortex core and gives rise to a circulating flow. Each vortex carries a single quantum of circulation $\kappa = 2\pi\hbar/m$ where $m$ is the mass of the constituent atoms or atom-pairs (for superfluid $^3$He-B, $\kappa = \pi\hbar/2m_3=0.662\times10^{-7}\,{\rm m^2/s}$, where $m_3$ is the mass of a $^3$He atom).

At low temperatures, quantum vortices move with the local superfluid velocity \cite{BDV}. Instabilities and reconnections of quantized vortices result in the formation of a vortex tangle which displays complex dynamics,  as each vortex moves according to the collective velocity field of all other vortices \cite{Donnelly-book,BDV-book}. This complex, disordered flow is known as quantum turbulence. A property that characterizes the intensity of quantum turbulence is the vortex line density, $L$ (${\rm m}^{-2}$), that is, the total length of vortex lines per unit volume. At higher temperatures the thermal excitations form a normal fluid component which may couple to the superfluid via mutual friction. The resulting behavior is quite complex, depending on whether the normal fluid also becomes turbulent. Here we only consider the low temperature regime in $^3$He-B, $T\lesssim0.3T_c$, where $T_c\approx0.9\,{\rm mK}$, is the critical temperature of superfluid transition at zero pressure. In this temperature regime, thermal excitations no longer form the normal fluid, but a few remaining excitations, whose mean free paths greatly exceed the typical size of the experimental cell, form a ballistic gas which has no influence on the vortex dynamics.

Even in this low temperature case, the properties of quantum turbulence are not yet fully understood; besides, it is interesting to compare the properties of pure quantum turbulence with those of classical turbulence. Understanding their relationship may eventually lead to a better understanding of turbulence in general.

During the last two decades there has been significant progress in the experimental studies of quantum turbulence. Experimental techniques are being developed to measure properties of quantum turbulence at low temperatures, both in $^4$He (see e.g. Refs.~\cite{Golov,Walmsley-2008,Walmsley-2014,Zmeev2015,Yano,Nichol,McClintock}) and in $^3$He-B \cite{Fisher,Bradley2,Bradley3,Finne,Hosio}. The most promising technique which is being developed for superfluid $^3$He-B utilizes the Andreev reflection of quasiparticle excitations from superfluid flow~\cite{Fisher_review}. (This technique was first developed to study quantum turbulence in stationary samples~\cite{Fisher,Bradley2,Bradley3}, and has since been extended to rotating samples~\cite{Hosio}.)

Andreev reflection arises in a Fermi superfluid as follows. The energy-momentum dispersion curve $E(\pp)$ for excitations has a minimum at the Fermi momentum, $\pF$, corresponding to the Cooper pair binding energy $\Delta$. Quasiparticle excitations have $|\pp| > \pF$ whereas quasiholes have $|\pp|< \pF$. On moving from one side of the minimum to the other, the excitation's group velocity reverses: quasiholes and quasiparticles with similar momenta travel in opposite directions. In the reference frame of a superfluid moving with velocity $\vv$, the dispersion curve tilts to become $E(\pp)+\pp\cdot\vv$ \cite{Fisher_review}. Quasiparticles which move into a region where the superfluid is flowing along their momentum direction will experience a potential barrier. Quasiparticles with insufficient energy are reflected as quasiholes. The momentum transfer is very small so there is very little force exerted whilst the outgoing quasiholes almost exactly retrace the path of the incoming quasiparticles~\cite{BSS1}. Andreev reflection therefore offers an ideal passive probe for observing vortices at very low temperatures.

Numerical simulations of Andreev reflection for two-dimensional vortex configurations have shown that the total reflection from a configuration of vortices may be significantly less than the sum of reflections from individual vortices~\cite{BSS2,BSS3}. This phenomenon is called “partial screening”. The Andreev scattering cross-sections of three-dimensional isolated vortex rings and the role of screening effects from various configurations of vortex rings in $^3$He-B were studied in Ref.~\cite{BSS4}.

In this work we analyze numerically the Andreev reflection of quasiparticles by three-dimensional vortex tangles in $^3$He-B. We use moderate tangle densities which are comparable to those typically observed in the experiments. The plan of the paper is as follows. In Sec.~\ref{sec:theornumer} we introduce the basic equations governing the dynamics of the vortex tangle and the associated flow field, the equations governing the propagation of thermal quasiparticle excitations, and the method for the calculation of the Andreev-reflected fraction of quasiparticle excitations incident on the vortex tangle. In Sec.~\ref{sec:vortgeneration} we describe our method for the numerical generation of vortex tangles and discuss their properties. In Sec.~\ref{sec:quasiclassical} we present our results for the simulation of the Andreev reflection from saturated, quasiclassical vortex tangles and analyze the dependence of the total reflection coefficient and the reflection scattering length on the vortex line density.  In subsection~\ref{sec:exptl} we describe the experimental techniques and the results for the experimental measurements of the Andreev reflection from quantum turbulence for comparison with our simulations.

In Sec.~\ref{sec:ultraquantum} we compare these results with those obtained for the Andreev reflection from unstructured, `ultraquantum' turbulence. In Sec.~\ref{sec:screening} we discuss the mechanisms of screening in the Andreev reflection from vortex tangles.  In Sec.~\ref{sec:fluctuations} we present our numerical results for the spectral properties of the Andreev reflection and of the vortex line density, analyze the correlation between these two statistical properties and compare them with the statistical properties of the experimentally-observed Andreev-retroreflected signal. We discuss our findings and their interpretation in Sec.~\ref{sec:interpretations}.

\section{Dynamics of the vortex tangle and propagation of thermal quasiparticle excitations: governing equations and numerical methods}
\label{sec:theornumer}

Since our spatial resolution scale is much larger than the coherence length, we will model the vortex filaments as infinitesimally thin vortex lines. At low temperatures, $T\lesssim0.3T_c$ such that mutual friction can be ignored, the superflow field $\vv(\rr,\,t)$ and the dynamics of the vortex tangle are governed by the coupled equations~\cite{Saffman}

\begin{equation}
\vv(\sss,\,t)=-\frac{\kappa}{4\pi}\oint\limits_{\cal L}
\frac{(\sss-\rr)}{\vert\sss-\rr\vert^3}\times d\rr\,,
\quad \frac{d\sss}{dt}=\vv(\sss,\,t)\,,
\label{eq:biot_savart}
\end{equation}
where the Biot-Savart line integral extends over the entire vortex configuration $\cal L$, with $\sss=\sss(t)$ identifying a point on a vortex line. In the first equation, the singularity at $\rr=\sss$ is removed in a standard way~\cite{BagBar}. The second, kinematic equation states that the vortex line moves with the local superfluid velocity. This is valid at low temperatures where mutual friction is absent~\cite{BDV-book}. Our numerical simulation of Eqs.~(\ref{eq:biot_savart}) is made in a periodic cubic box for vortex tangles of line densities up to $L \sim 10^8\,{\rm m^{-2}}$, see Sec.~\ref{sec:vortgeneration} for detail.

In Ref.~\cite{BSS1}, based on the approach developed by Greaves and Leggett~\cite{Leggett}, we derived the following semiclassical Hamiltonian equations governing the propagation of thermal quasiparticle excitations in $^3$He-B:
\begin{equation}
\frac{d\rr}{dt}=\frac{\ep}{\sqrt{\ep^2+\Delta^2}}\frac{\pp}{m^*}+\vv, \quad \frac{d\pp}{dt}=-\frac{\partial}{\partial\rr}(\pp\cdot\vv)\,,
\label{eq:motion}
\end{equation}
where $\rr(t)$ represents the position of a quasiparticle excitation taken as a compact object, $\pp(t)$ is the momentum of excitation, $\epsilon_p=p^2/(2m^*)-\epsilon_F$ is the ``kinetic'' energy of a thermal excitation relative to the Fermi energy, $m^*\approx3.01m_3$ is the effective mass of an excitation, with $m_3$ being the bare mass of a $^3$He atom, and $\Delta$ is the superfluid energy gap. Equations~(\ref{eq:motion}) were derived by considering the propagation of excitations at distances from the vortex core large compared to the coherence length (at zero temperature and pressure the coherence length is $\xi_0\approx50\,{\rm nm}$).

In principle, interactions of thermal excitations with the vortex tangle can be analyzed by simultaneously solving Eqs.~(\ref{eq:motion}), governing the propagation of quasiparticles, and Eqs.~(\ref{eq:biot_savart}), governing the dynamics and the superflow field of the tangle. This is exactly how some of the authors of this paper earlier analyzed the Andreev reflection from two-dimensional vortex configurations~\cite{BSS2,BSS3} and three-dimensional quantized vortex rings~\cite{BSS4}. However, Eqs.~(\ref{eq:motion}) are stiff, and solving them simultaneously with Eqs.~(\ref{eq:biot_savart}) for saturated three-dimensional vortex tangles requires considerable computing time and resources. For tangles of moderate line densities (up to $L \sim 10^8\,{\rm m}^{-2}$) considered in this work, an alternative, much simpler method for evaluating the  flux of quasiparticles Andreev-reflected by the tangle can be developed based on the properties~\cite{BSS1} of solution of Eqs.~(\ref{eq:motion}) and on the arguments~\cite{PNAS.111.4659.Fisher,PRL.115.015302.Baggaley} formulated below.

The alternative method makes use of the facts~\cite{BSS1}; first, that at moderate tangle densities, the Andreev reflected excitation practically retraces the trajectory of the incident quasiparticle, and secondly that the time scale of propagation of the quasiparticle excitation within the computational box (of 1~mm size in our simulations) is much shorter than the characteristic timescale, $\tau_f\sim1/(\kappa L)$ of the fluid motion in the tangle. In analyzing the quasiparticle trajectories, these results allow us to assume a `frozen' configuration of vortices and the associated flow field.

Assuming that the statistically uniform and isotropic vortex tangle occupies the cubic computational box of size $D$, consider the propagation of a flux of quasiparticles at normal incidence on one side of the box in (say) the $x$-direction. Also assume that the quasiparticle flux is uniformly distributed over the $(y,\,z)$ plane and covers the full cross section of the computational box.

Ignoring angular factors, the incident flux of quasiparticles, as a function of the position $(y,\,z)$, can be written as~\cite{PNAS.111.4659.Fisher}
\begin{equation}
\langle nv_g\rangle^i_{(y,z)}=\int^{\infty}_{\Delta}g(E)f(E)v_g(E)dE\,,
\label{eq:Eflux}
\end{equation}
where $g(E)$ is the density of states, and $f(E)$ is the Fermi distribution function which we can approximate at low temperatures to the Boltzmann distribution, $f(E) = \exp(-E/k_BT)$. The average energy of a quasiparticle in this flux is $\langle E\rangle = \Delta + k_BT$.

Since the typical quasiparticle group velocity is large compared to a typical superfluid flow velocity $v$, the integral is simplified by noting that $g(E)v_g(E) = g(p_F)$ is the constant density of momentum states at the Fermi surface. Equation~(\ref{eq:Eflux}) therefore reduces to
\begin{align}
\begin{split}
\langle nv_g\rangle^i_{y,z}&=g(\pF)\int^{\infty}_{\Delta}\exp(-E/k_BT)\,dE \\
&= g(p_F) k_BT\exp(-\Delta/k_BT)\,.
\end{split}
\label{eq:Exfluxsimp}
\end{align}

In the presence of a tangle, and hence regions of varying superfluid velocity $\vv$, a quasiparticle moving into a region where the superfluid is flowing along their momentum direction experiences a force $\dot{\pp}=-\nabla(\pp\cdot\vv)$, which decreases its momentum towards $p_F$ and reduces its group velocity. If the flow is sufficiently large the quasiparticle is pushed around the minimum, becoming a quasihole with a reversed group velocity. Consequently, the flux of quasiparticles transmitted through a tangle, $\langle nv_g\rangle^t$ is calculated by the integral~(\ref{eq:Exfluxsimp}) in which the lower limit is replaced by $\Delta+\max(\pp\cdot\vv)\approx\Delta+\pF v_x^{\max}$, where $v_x^{\max}$ is the maximum value of the $x$-component of superfluid velocity, found from the solution of Eqs.~(\ref{eq:biot_savart}), along the quasiparticle's rectilinear trajectory [that is, for fixed $(y,\,z)$]:
\begin{align}
\begin{split}
\langle nv_g\rangle^t_{y,z} = g(p_F)\int^{\infty}_{\Delta + p_Fv_x^{\max}}\exp(-E/k_BT)\,dE \\
= g(p_F)k_BT\exp[-(\Delta + p_Fv_x^{\max})/k_BT]\,.
\end{split}
\label{eq:Exfluxsimp2}
\end{align}

For the position $(y,\,z)$ on the face of computational box, the Andreev reflected fraction of quasiparticles incident on a tangle along the $x$-direction is, therefore,
\begin{equation}
f_{y,z} = 1 - \frac{\langle nv_g\rangle^t_{y,z}}{\langle nv_g\rangle^i_{y,z}} = 1 - \exp\left[-\frac{p_Fv_x^{\max}}{k_BT}\right]\,.
\label{eq:ExAndsimp}
\end{equation}
The total Andreev reflection $f_x$ is the average of the Andreev reflections for all positions of the $(y,\,z)$ plane. The same calculation is repeated for the flux of quasiholes. The results are then combined to yield the reflection coefficient for the full beam of thermal excitations.

In our simulation, the initial positions of quasiparticle excitations on the $(y,\,z)$ plane were uniformly distributed with spatial resolution $\approx6\times6\,{\rm \mu m}^2$. More details of numerical simulation will be given below in Sec.~\ref{sec:vortgeneration}.

In the following sections, to directly compare our simulations with experimental observations, we fix the temperature at $T=0.15T_c$ and use parameters appropriate to $^3$He-B at low pressure: Fermi energy $\eF=2.27\times10^{-23}\,{\rm J}$, Fermi momentum $\pF = 8.28 \times 10^{-25}\,{\rm kg\,m\,s}^{-1}$, superfluid energy gap $\Delta_0=1.76k_BT_c$, and quasiparticle effective mass $m^*\approx 3.01\times m_3=1.51 \times10^{-26}\,\rm kg$.

\section{Numerical generation and properties of vortex tangles}
\label{sec:vortgeneration}

\begin{figure}[t]
\begin{center}
    \includegraphics[width = 0.96\linewidth]{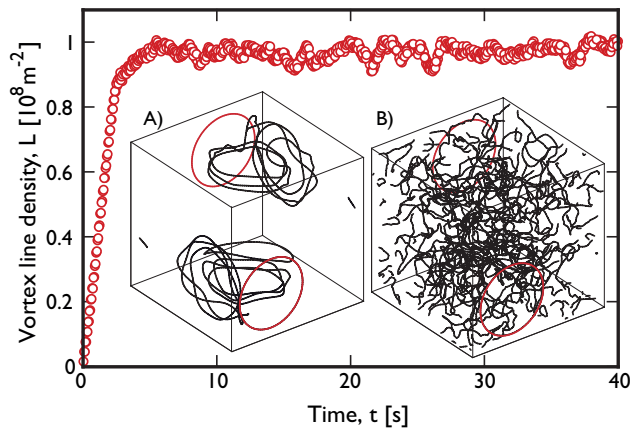}
    \caption{Evolution of the vortex line density of the tangle generated by injection of vortex rings. Insets (A) and (B) demonstrate snapshots of the tangle at the moment of ring injection from the opposite corners of the computational domain  for initial and steady state of simulation correspondingly (see text for details).}
    \label{fig:transient}
\end{center}
\end{figure}

Motivated by experimental studies~\cite{Bradley4}, we numerically simulate the evolution of a vortex tangle driven by vortex loop injection\cite{PRL.115.015302.Baggaley}. Two rings of radius $R_i=240\,\mu{\rm m}$ are injected at opposite corners of the numerical domain (see Fig.~\ref{fig:transient}(A), where the two injected vortex rings, marked in red, can be seen entering from the left and right sides of the box repeatedly with a frequency $f_i=10\,{\rm Hz}$. The vortex loops injected into the simulation box collide and reconnect to rapidly generate a vortex tangle. The time evolution of the vortex tangle and the superfluid velocity $\vv(\rr,\,t)$ were simulated by means of the vortex filament method, that is by solving Eqs.~(\ref{eq:biot_savart}), in a cubic box of size $D = 1\,\rm{mm}$ with periodic boundary conditions~\cite{BagBar}. (As can be seen from Eq.~(\ref{eq:ExAndsimp}), the accuracy of simulation of the superflow field is crucially important for the calculation of the Andreev reflection.) To ensure good isotropy, the plane of injection of the loops is switched at both corners with a frequency of $f_s=3.3\,{\rm Hz}$. We integrate Eqs.~(\ref{eq:biot_savart})  for a period of 380~s.

For dense vortex tangles, we use the tree-method with a critical opening angle of 0.4 rad~\cite{BagBar}. For small line densities, of the order of few ${\rm mm}^{-2}$ or less, since the tree-method loses accuracy, calculations were performed instead by the direct evaluation of the Biot-Savart integrals. The evolution of the vortex line density for the first 40~s of the simulation is shown in Fig.~\ref{fig:transient}.

\begin{figure}[ht]
    \begin{center}
    \includegraphics[width = 0.96\linewidth]{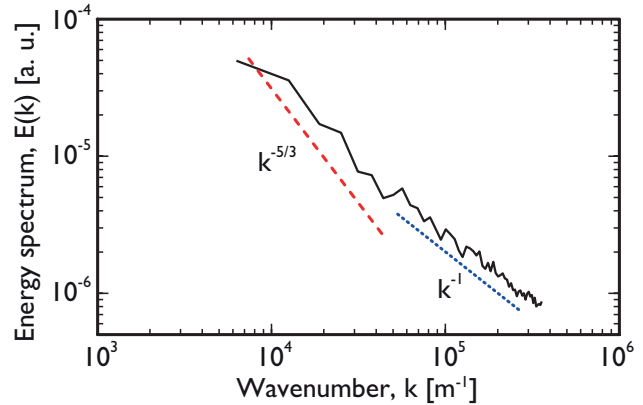}
    \caption{The energy spectrum\cite{PRL.115.015302.Baggaley}, $E(k)$ of the steady state vortex tangle vs the wavenumber $k$. The slope of the spectrum is consistent with the $k^{-5/3}$ Kolmogorov scaling (red dashed line) for small wavenumbers  and $k^{-1}$ scaling (blue dotted line) for large wavenumbers.}
    \label{fig:tanglspectr}
\end{center}
\end{figure}

After an initial transient, lasting for about 5~s, the energy injected into the box is balanced by the numerical dissipation, which arises from the spatial resolution ($\approx$6~$\mu$m) of the numerical routine (small scale structures such as small vortex loops and high frequency Kelvin waves of wavelength shorter than this spatial resolution are therefore lost), and the tangle saturates to reach the equilibrium state. In this statistically steady state the vortex line density, $L(t)$, fluctuates around its equilibrium value $\langle L\rangle = 9.7 \times 10^7$ m$^{-2}$ which corresponds to the characteristic intervortex distance of $\ell\approx \langle L\rangle^{-1/2} = 102$~$\mu$m. A snapshot of this tangle is shown in the inset (B) of Fig.~\ref{fig:transient}.
In Fig.~\ref{fig:tanglspectr} we show the energy spectrum, $E(k)$ of this
tangle (here $k$ is the wavenumber). The energy spectrum, representing the distribution of kinetic energy over the length scales, is an important property of homogeneous isotropic turbulence. In the statistically steady state the spectrum is consistent, at moderately large scales ($k_D < k < k_\ell$, where $k_D = 2\pi/D \approx 6 \times 10^3$~m$^{-1}$ and $k_\ell = 2\pi/\ell \approx 6 \times 10^4$ m$^{-1}$), with the Kolmogorov $-5/3$ spectrum shown in Fig.~\ref{fig:tanglspectr} by the red dashed line. In classical turbulence the Kolmogorov $-5/3$ scaling is the signature of a `cascade' mechanism which transfers energy from large length scales to small length scales. The blue dotted line shows the $k^{-1}$ power spectrum expected for small scales ($k>k_\ell$) arising from the superfluid flow around individual vortex cores. Fig.~\ref{fig:tanglspectr} also shows that most of the energy of the turbulence is at the smallest wavenumbers (the largest length scales). This property is caused by  partial polarization of the vortex tangle: parallel vortex lines locally come together, creating metastable vortex bundles, whose net rotation represents relatively intense large-scale flows which are responsible~\cite{BagLauBar} for the pile-up of energy near $k \approx k_D$.

\section{Andreev reflection from a quasiclassical tangle}
\label{sec:quasiclassical}

\subsection{Two-dimensional map of local Andreev reflections}
\label{sec:2Dmap}

\begin{figure}[t]
    \begin{center}
    \includegraphics[width=0.95\linewidth]{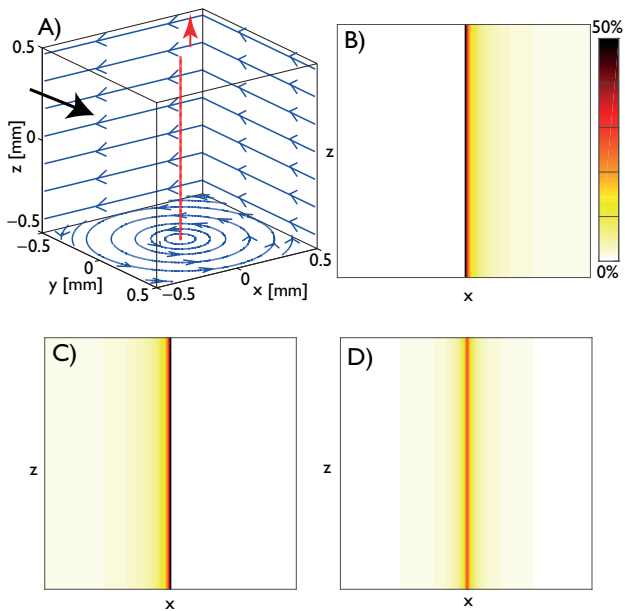}
    \caption{Representation of the reflection coefficient for excitations incident on the flow field of an isolated, rectilinear vortex line. (A) Velocity flow field around a rectilinear vortex. The red arrow shows the direction of vorticity. The black arrow indicated the direction of injection of quasiparticles and quasiholes. (B) and (C) show the quasiparticle and quasihole reflection, respectively. (D) The total reflection of incoming excitation flux. Darker (red/yellow) regions correspond to the higher reflection coefficient.}
    \label{fig:ARstraight}
    \end{center}
\end{figure}

We start with illustrating the numerical calculation, based on the method described above in Sec.~\ref{sec:theornumer}, of the Andreev reflection from an isolated, rectilinear vortex shown in Fig.~\ref{fig:ARstraight}(A). The vortex's circulating flow velocity is $v(r)=\kappa/(2\pi r)$, where $r$ is the distance from the vortex core (for theoretical analysis and analytical calculation of Andreev reflection from the rectilinear vortex see Refs.~\cite{BSS1,PNAS.111.4659.Fisher,Sergeev2015}).

The velocity field around the vortex was computed using Biot-Savart integrals. To ensure the homogeneity of the velocity flow profile across the simulation volume, the actual vortex extends another four lengths above and below the cube shown in the figure.

We calculate the Andreev reflection of a beam of excitations which propagate in the $y$-direction orthogonally to the $(x,\,z)$ face of the computational cube (the black arrow shows direction of the incoming excitation). This calculation gives a measure of the Andreev reflection as a two-dimensional contour map across the full cross section of the incident quasiparticle beam. In Fig.~\ref{fig:ARstraight}, panels (B) and (C) show, respectively, the reflection of quasiparticles ($p>\pF$) and quasiholes ($p<\pF$) from the rectilinear, quantized vortex that has a direction of vorticity pointing up. The darker (red/yellow) regions corresponding to the higher reflectivity clearly show the ``Andreev shadows'' and are located symmetrically on the immediate right of the vortex line for quasiparticles and on the immediate left for quasiholes. Note also that for a `real' thermal beam consisting of excitations of both sorts the Andreev shadow of a single vortex should be symmetric with respect to the vortex line and can be imagined as a superposition of panels~(B) and (C). The panel (D) illustrates the total Andreev reflection from the rectilinear vortex.

We use the same method of evaluating Andreev reflection for our simulated tangle. The velocity field of the vortex tangle was computed using a slightly modified tree-method with a critical opening angle of 0.2, compared to the previously published result~\cite{PRL.115.015302.Baggaley} where a critical opening angle of 0.4 was used. The modified tree-method algorithm continues propagation along the tree when the desired critical opening angle is reached but the local tangle polarization (defined in Refs.~\cite{Lvov-PRB-2007,Baggaley-PoF-2012}) exceeds 3.8. The current simulation shows a slightly better agreement with the direct Biot-Savart calculation, but overall results obtained using the modified tree-method replicate previous conclusions.
\begin{figure}[ht]
    \begin{center}
    \includegraphics[width=0.95\linewidth]{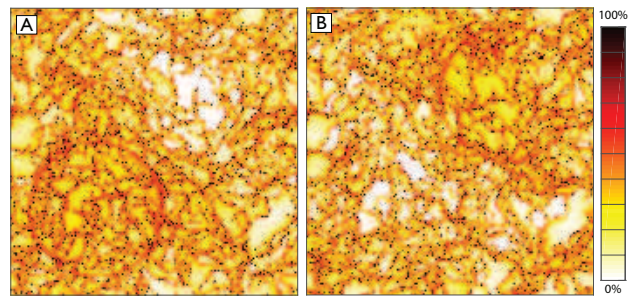}
    \caption{Two-dimensional representation of the reflection coefficient for excitations incident on the $(x,\,z)$ plane of the computational cubic domain. The vortex cores are visible as thin dark lines. (A):~reflection of quasiparticles ($p>\pF$); the extended darker and brighter regions correspond to the large scale flow into and out of the page, respectively. (B):~reflection of quasiholes ($p<\pF$); darker and brighter regions indicate the large scale flow in the directions opposite to those in panel~(A).}
    \label{fig:2Dmap}
    \end{center}
\end{figure}

Figure~\ref{fig:2Dmap} shows a distribution of the local Andreev reflection coefficient $f_{x,z}$ [calculated using Eq.~(\ref{eq:ExAndsimp})] on the $(x,\,z)$ plane for a tangle with the saturated equilibrium line density $\langle L\rangle=9.7\times10^7\,{\rm m}^{-2}$ (Fig.\ref{fig:transient} inset~(B)). The darker (red/yellow) regions correspond to the higher Andreev reflection; clearly visible, very dark, thin ``threads'' correspond to a high/complete Andreev reflection which is expected in the vicinity of vortex cores. The overall distribution of dark threads corresponding to vortex cores is roughly uniform and indicates that our simulated turbulent tangle is practically homogeneous. Panel~(A) illustrates the reflection of quasiparticle excitations with $p > p_F$, and (B) shows the reflection of quasiholes ($p < p_F$). The shadows of large vortex rings instantaneously injected into the computational domain to sustain the statistically steady state are also visible.

The individual maps of Andreev reflection for two species of excitations [quasiparticles in Fig.~\ref{fig:2Dmap}~(A) and quasiholes in Fig.~\ref{fig:2Dmap}~(B)] show the presence of large scale flow, consistent with the quasiclassical (Kolmogorov) character of the simulated vortex tangle whose energy spectrum scales as $k^{-5/3}$ in the ``inertial'' range of moderately small wavenumbers, see Fig.~\ref{fig:tanglspectr}. The extended darker regions in Fig.~\ref{fig:2Dmap}~(A) correspond to large scale flows whose $y$-component of velocity is directed into the page; the brighter regions show large scale flows directed out of the page. For quasiholes reflection shown in panel~(B) the meaning of darker and brighter regions is opposite to that of the panel~(A). Consequent direct calculation of the superflow patterns within the simulation volume has confirmed the presence of large scale flows in the corresponding regions.

We should emphasize here that in the numerical experiment described above the \textit{direction} of the large scale flow could be determined only because the separate simulations, for the same tangle, were performed for two different species of excitations, that is for quasiparticles with $p>\pF$ and quasiholes with $p<\pF$. In the real physical experiment the thermal beam consists of excitations of both sorts whose separation is, of course, not possible. For such a beam the Andreev reflection should be the same both for receding and approaching flows. While the presence of large scale flow may still be detected by the enhancement of the Andreev reflection, an identification of its direction would hardly be possible.

A two-dimensional representation of the reflection coefficient for the \textit{full} beam of excitations Andreev-reflected by the simulated tangle is illustrated below in Fig.~\ref{fig:ARtransient} by the insets (A) and (B) for the initial and steady state of simulation, respectively.

\subsection{Total Andreev reflection vs vortex line density}
\label{sec:total}

We now turn to the results of our calculation of the total Andreev reflection from the vortex tangle. Our aim is to find the total reflection coefficient, $f_R$, that is, the fraction of incident excitations Andreev-reflected by the tangle, as a function of the vortex line density. Here the total reflection coefficient is defined as $f_R=(f_x+f_y+f_z)/3$, where $f_x$, $f_y$, and $f_z$ are the total reflection coefficients of the excitation beams incident, respectively, in the $x$, $y$, and $z$ directions on the $(y,\,z)$, $(x,\,z)$, and $(x,\,y)$ planes of the cubic computational domain.

\begin{figure}[b]
    \begin{center}
        \includegraphics[width=0.95\linewidth]{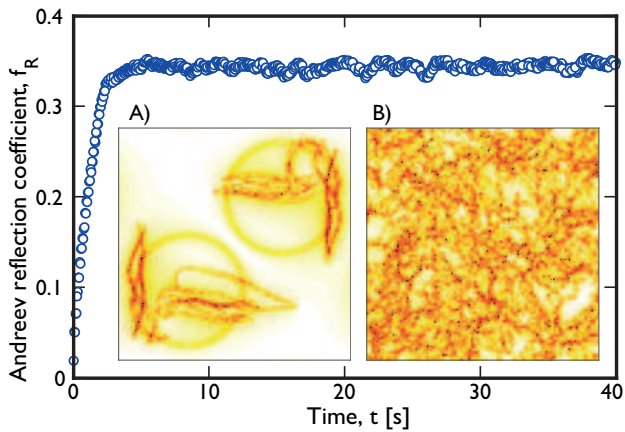}
        \caption{Evolution of the total Andreev reflection for the tangle generated by injection of vortex rings (see text for details). Insets (A) and (B) show a two-dimensional representation for the full beam of excitations for the initial and steady state of simulation, respectively.}
        \label{fig:ARtransient}
    \end{center}
\end{figure}

\begin{figure}[t]
    \begin{center}
    \includegraphics[width=7.5cm]{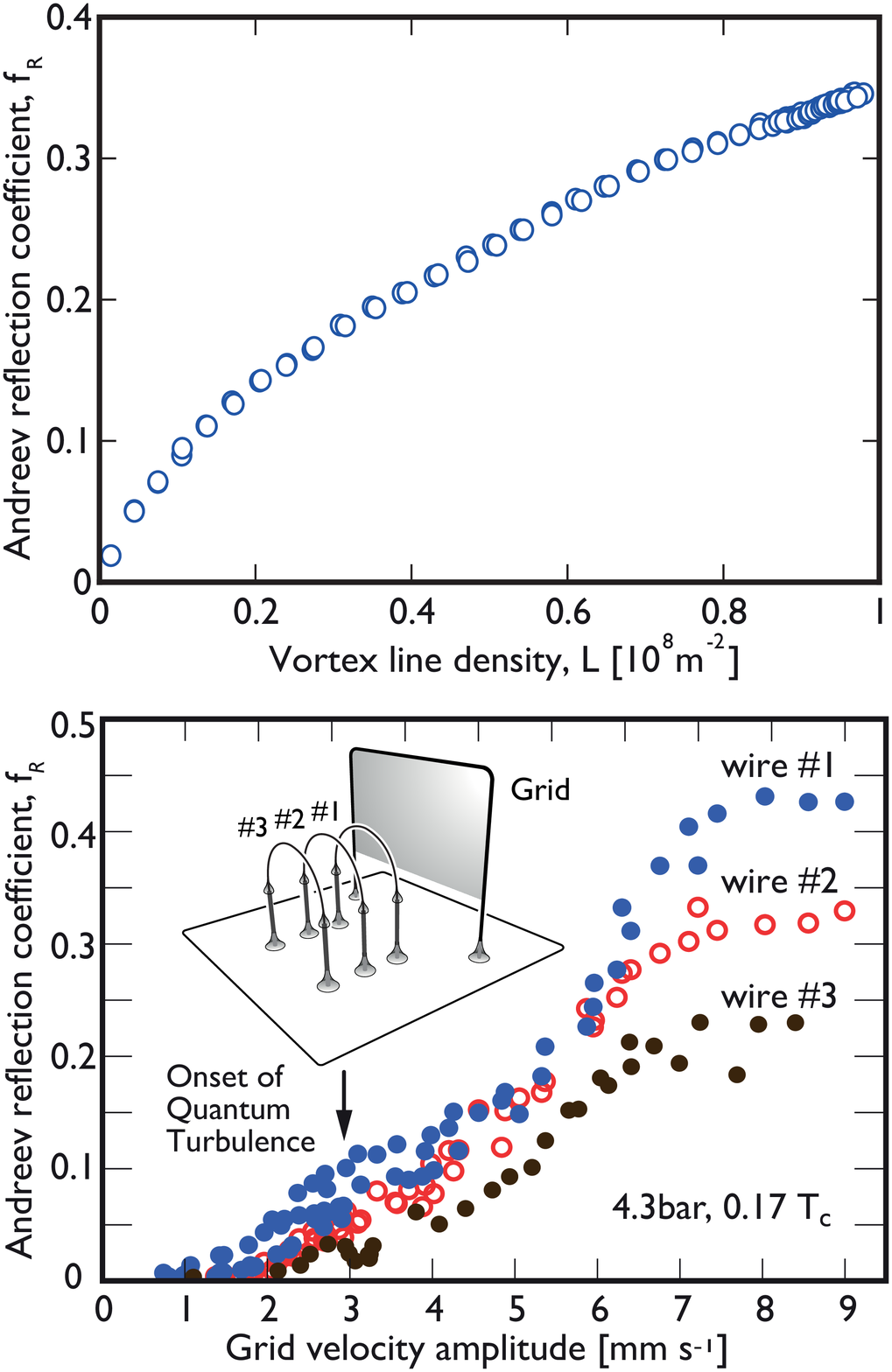}
    \caption{Top: the total reflection coefficient obtained from the numerical simulations, plotted against the line density of the vortex tangle. Bottom: experimental measurements of the fraction of Andreev scattering from vortices generated by a grid plotted versus grid's velocity\cite{PRL.115.015302.Baggaley}. Measurements are shown for three vibrating wires at different distances from the grid as shown in the inset, see text.}
    \label{fig:ARvsL}
    \end{center}
\end{figure}

For the tangle numerically generated by injected rings of radius $R_i=240\,\mu{\rm m}$ with the injection and switching frequencies $f_i=10\,{\rm Hz}$ and $f_i=3.3\,{\rm Hz}$, respectively (see above Sec.~\ref{sec:vortgeneration} for details), we monitor simultaneously the evolution of the vortex line density (see Fig.~\ref{fig:transient}) and of the reflection coefficient calculated using the method described in Sec.~\ref{sec:theornumer} [see Eq.~(\ref{eq:ExAndsimp}) and the text that follows] during the transient period~\cite{timescale} and for some time after the tangle has saturated at the statistically steady-state, equilibrium value of the vortex line density, $\langle L\rangle = 9.7 \times 10^7$ m$^{-2}$, see Fig.~\ref{fig:ARtransient}. After a very short initial period of less than 1~s, the vortex configuration becomes, and remains during the transient and afterwards practically homogeneous and isotropic. This enables us to re-interpret the results shown in Figs.~\ref{fig:transient} and \ref{fig:ARtransient} in the form of the Andreev-reflected fraction, $f_R$ as a function of the vortex line density. This function~\cite{fvsL} is shown by the top panel of Fig.~\ref{fig:ARvsL}.

As can be seen from Fig.~\ref{fig:ARvsL}, at small line densities, $L\lesssim2\times10^7\,{\rm m}^{-2}$ the reflection coefficient rises rapidly and linearly with $L$. During the initial stage of the generation of the tangle, the injected rings form a dilute gas and are virtually non-interacting. As the computation time advances, more rings enter the simulation box, start to interact, collide, reconnect, and eventually form a tangle which absorbs all injected rings. However, at small times, even after the tangle has been formed, the vortex line density remains relatively small so that the average distance between vortex lines is relatively large. In this case the flow around each vortex is practically the same as the flow field of a single, isolated vortex~\cite{smallL}. In a sufficiently dilute vortex configuration a screening of Andreev reflection (discussed below in Sec.~\ref{sec:screening}) can also be neglected so that the total reflection is a sum of reflections from individual vortex lines; this explains a linear growth with $L$ of the reflection coefficient at low line densities.

\subsection{Comparison with experiment}
\label{sec:exptl}

We compare the simulation with the experiment~\cite{PNAS.111.4659.Fisher,Bradley4} which has the configuration similar to that shown in the inset of the lower panel of Fig.~\ref{fig:ARvsL}.  The vorticity is produced by the oscillating grid and surrounds the vibrating wire detectors.  The tangle's flow field Andreev-reflects ambient thermal excitations arriving from ``infinity'', shielding the wires and reducing the damping. The reduction in damping on each wire (placed at 1.5\,mm,  2.4\,mm and 3.5\,mm from the grid) provides the measure of the Andreev reflection by the tangle, see Ref.~\cite{PNAS.111.4659.Fisher,PRL.115.015302.Baggaley}. The lower panel of Fig.~\ref{fig:ARvsL} shows the fractional reflection of quasiparticles incident on each wire.

The numerical and experimental data plots in Fig.~\ref{fig:ARvsL} have similar shapes. While the vortex line density $L$ cannot be obtained directly from the measurements, we expect the local line density of the quantum turbulence to increase steadily with increasing grid velocity. However, the onset of turbulence is rather different in the simulations and in the experiment. In the simulations, approaching injected vortex ring pairs are guaranteed to collide and form a tangle, whereas the vibrating grid emits only outward-going vortex rings, with the ring flux increasing steadily with increasing grid velocity. At low grid velocities, rings propagate ballistically with few collisions \cite{Bradley2,FujiYamaPRB}. At higher velocities, ring collisions increase giving rise to the vortex tangle. For the data of Fig.~\ref{fig:ARvsL}, this occurs when the  grid velocity exceeds $\sim$ 3\,mm/s. The data for lower velocities correspond to reflection from ballistic vortex rings and can be ignored for the current comparison.

At higher grid velocities/tangle densities, the fraction of Andreev-reflected excitations rises at an increasing rate, finally reaching a plateau. The plateau region is prominent in the experiments when the grid reaches velocities approaching a third of the Landau critical velocity~\cite{LambertPhB,PRL.69.1073.Fisher.1992} and may be an artifact of data analysis model that ignores a local flux of quasiparticles produced by pair-breaking.

For the wire closest to the grid, the experimentally observed absolute value of the reflectivity is almost identical to that obtained from our simulation. This excellent agreement is perhaps fortuitous given that in the experiments quasiparticles travel through 1.5--2.5\,mm of turbulence to reach the wire, compared with 1\,mm in the simulation, thus larger screening might be expected. A better comparison will require high-resolution experiments to separate the effects caused by variations of tangle density from those caused by the increase of quasiparticle emission by the grid.

\subsection{The Andreev scattering length}
\label{sec:length}

To compare the amount of Andreev scattering from vortex tangles with that of isolated vortices, it might be useful to define an Andreev scattering length. In Ref.~\cite{BSS4} the cross section, $\sigma$, for Andreev scattering was defined as
\begin{equation}
\sigma= \frac{\dot{N}}{\langle n\vg\rangle},
\label{eq:AndScat}
\end{equation}
where $\dot{N}$ is the number of excitations reflected per unit time, and $\langle n\vg\rangle$ is the number flux of incident excitations. From the cross section $\sigma$ we define the Andreev scattering length, $b$, as
\begin{equation}
\sigma= b \mathcal{L},
\label{eq:AndScatlen}
\end{equation}
where $\mathcal{L}$ is the total length of vortex line within the computational box or the experimental cell. For the simulation box with side length $D$, $\mathcal{L} = LD^3$. From definition~(\ref{eq:AndScat}) of the cross section, the total reflectivity of a vortex configuration is $f = \sigma/A$, where $A$ is the cross sectional area of the incident excitation beam. For our simulations, the Andreev scattering length is thus given by
\begin{equation}
b = f/(DL)\,.
\label{eq:AndScatlensim}
\end{equation}

\begin{figure}[b]
    \begin{center}
    \includegraphics[width=0.95\linewidth]{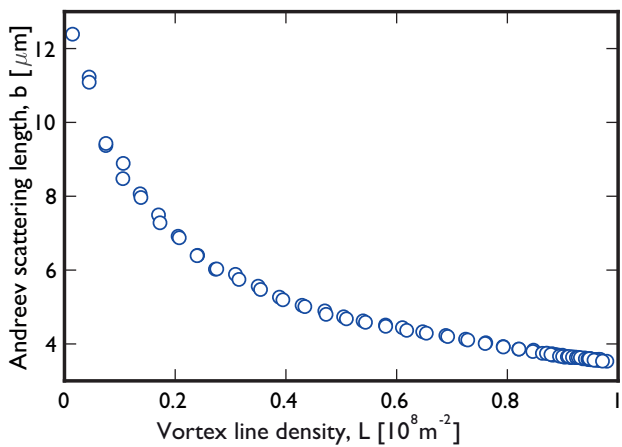}
    \caption{Andreev scattering length, $b$, vs  vortex line density, $L$.}
    \label{fig:s-length}
    \end{center}
\end{figure}

The Andreev scattering length in the vortex tangles, obtained from Eq.~(\ref{eq:AndScatlensim}) and our numerical simulation, is plotted in Fig.~\ref{fig:s-length}.
For the isolated, rectilinear quantized vortex line, the theoretical analysis and analytical calculation of the Andreev scattering length were given in Ref.~\cite{Sergeev2015}. A simpler estimate for the \textit{scale} of Andreev scattering length, used for the initial interpretation of experiments~\cite{Fisher,Bradley2,Bradley3,Bradley4}, can be obtained as follows. Consider quasiparticle excitations incident on a straight vortex line perpendicular to the excitation flux. On one side of the vortex, an incident quasiparticle will experience a potential barrier $p_F v = \hbar/(2m_3r)$, where $r$ is the impact parameter. The average thermal quasiparticle energy is $\Delta+k_BT$. An approximate value for the Andreev scattering length $b$ is the impact parameter $b_0$ at which an excitation with the average energy is Andreev reflected. Equating the potential barrier with $k_BT$ yields
\begin{equation}
b_0 = \frac{p_F\hbar}{2m_3k_BT}\,.
\label{eq:AndImpact}
\end{equation}
For the parameters used in the simulations, appropriate to $^3$He-B at low pressure and $T = 150$~$\mu$K, estimate~(\ref{eq:AndImpact}) gives $b_0 = 4.2$~$\mu$m. This value is in a very good agreement with the Andreev scattering length of the vortex tangle at the highest line densities shown in Fig.~\ref{fig:s-length}. This validates the analysis previously used in Refs.~\cite{Fisher,Bradley2,Bradley3,Bradley4} to interpret the experimental data for $^3$He tangles.

We note, however, that while yielding the correct estimate for the scale~(\ref{eq:AndImpact}) of the Andreev scattering length, this simple model might be somewhat misleading since, even for an isolated vortex, a significant amount of Andreev reflection occurs at distances larger than $b_0$ from the vortex core, see Ref.~\cite{Sergeev2015} for details.

For dilute tangles, the Andreev scattering length is consistent with the value for large vortex rings, $b_{\rm rings} = 10$~$\mu$m, inferred from the results of numerical simulations of Ref.~\cite{BSS4}. It is clear that Andreev scattering from very dilute vortex tangles is dominated by reflection from individual vortices.

\section{Total reflection from the unstructured vortex tangle}
\label{sec:ultraquantum}

In the previous section we investigated Andreev reflection from quasiclassical quantum turbulence whose energy spectrum obeys the Kolmogorov $-5/3$ scaling law. However, there is also evidence for the existence of a second quantum turbulence regime, consisting of a tangle of randomly oriented vortex lines whose velocity fields tend to cancel each other out. This second form of turbulence, named `ultraquantum' or `Vinen' turbulence to distinguish it from the `quasiclassical' or `Kolmogorov' turbulence~\cite{Volovik,Skrbek2006}, has been identified both numerically~\cite{Baggaley-2012-ultra} and experimentally (in low temperature $^4$He~\cite{Walmsley-2008,Zmeev2015} and in $^3$He-B~\cite{Bradley3}), and (only numerically) in high temperatures $^4$He driven by a uniform heat current~\cite{Baggaley-Sherwin-2012,Sherwin-Baggaley-2015} (thermal counterflow). Physically, ultraquantum turbulence is a state of disorder without an energy cascade \cite{BSB2016}. Recently, it has been argued that ultraquantum turbulence has also been observed in the thermal quench of a Bose gas~\cite{Stagg2016}, and  in small trapped atomic condensates \cite{Cidrim2017}.

In the ultraquantum regime the vortex tangle is uncorrelated: There is no separation between scales (and hence no large-scale motion), and the only length scale is that of the mean intervortex distance, so that it can be expected that the total Andreev reflectivity of the ultraquantum tangle should differ from that of the quasiclassical quantum turbulence. However, we will demonstrate below that this is not the case.

To compare the Andreev reflection from two different types of quantum turbulence, we should, in principle, have generated numerically the ultraquantum vortex tangle whose line density would match that of the quasiclassical tangle described above in Secs.~\ref{sec:vortgeneration} and \ref{sec:quasiclassical}. However, it turns out that a numerical generation of the ultraquantum tangle with prescribed properties presents substantial difficulties.

\begin{figure}[t]
\begin{center}
\includegraphics[width=0.95\linewidth]{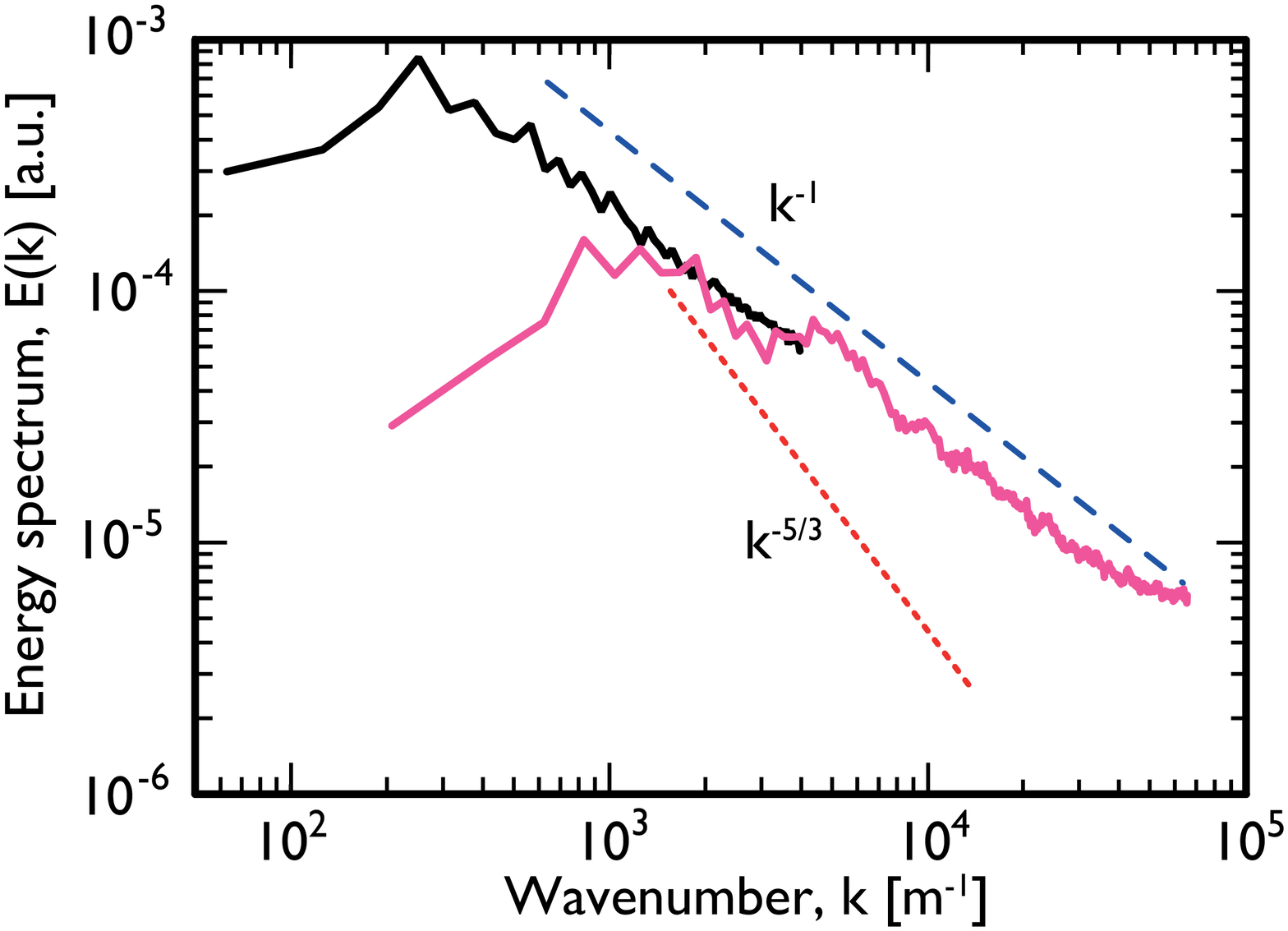}
\caption{Superfluid energy spectra (arbitrary units) obtained from numerical simulation. The black and red lines correspond to thermal counterflow~\cite{Sherwin-Baggaley-2015,BSB2016} and ultraquantum regime in the zero temperature limit~\cite{Baggaley-2012-ultra}, respectively. The red dotted and blue dashed lines are guides to the eye for the $k^{-5/3}$ and $k^{-1}$ scaling respectively.}
\label{fig:spectrum-ultraquantum}
\end{center}
\end{figure}

To overcome these difficulties, we will use an alternative approach. In Refs.~\cite{Baggaley-Sherwin-2012,Sherwin-Baggaley-2015,BSB2016} we numerically simulated  vortex tangles in $^4$He driven by a uniform normal fluid (to model thermally driven counterflow) and by synthetic turbulence in the normal fluid (to model mechanical driving, such as grids and propellers). Although these calculations model quantum turbulence in $^4$He at elevated temperatures ($1\,{\rm K}\lesssim T<T_\lambda$), the vortex tangle in the counterflow turbulence shares many features with the ultraquantum tangle in the zero temperature limit. Apart from small anisotropy in the streamwise vs transversal directions, the counterflow tangle is practically unstructured, with the intervortex distance being the only length scale, and the energy spectrum~\cite{Sherwin-Baggaley-2015,BSB2016} of the superfluid component, similar to that of the ultraquantum turbulence, nowhere conforms to the $-5/3$ scaling but has a broad peak at large wavenumbers, see Fig.~\ref{fig:spectrum-ultraquantum}, followed by the same $k^{-1}$ dependence numerically observed \cite{Baggaley-2012-ultra} in ultraquantum turbulence.

On the other hand, the tangle simulated for the mechanically driven turbulence~\cite{Baggaley-Sherwin-2012,Sherwin-Baggaley-2015} shares many features common with the quasiclassical quantum turbulence in the zero temperature regime. In particular, in both cases large-scale superflows are present, and the energy spectrum obeys Kolmogorov scaling for medium wavenumbers. Ignoring the normal fluid, the vortex tangles simulated for the thermally and mechanically driven turbulence in $^4$He can be regarded as modeling, at least qualitatively, the ultraquantum and quasiclassical regimes, respectively, of quantum turbulence in the low temperature $^3$He-B~\cite{KW-damping}.

In Refs.~\cite{Baggaley-Sherwin-2012,Sherwin-Baggaley-2015}, both the counterflow turbulence and the vortex tangle driven by the synthetic turbulence in the normal fluid were simulated starting from an arbitrary seeding initial condition and then letting $L$ grow and saturate to a statistically steady state. An important advantage is that in both cases the vortex line density was changing within the same interval from $L=0$ until saturation at $L\approx2.2\times10^8\,{\rm m}^{-2}$. For three realizations of the counterflow turbulence with different values of the counterflow velocity, and for three realizations of the mechanically driven turbulence with different values of the \textit{rms} normal fluid velocity (see Refs.~\cite{Baggaley-Sherwin-2012,Sherwin-Baggaley-2015} for details and values of parameters), we calculated the Andreev reflection from the unstructured (``ultraquantum'') and structured (``quasiclassical'') tangles. These calculations were performed for different values of the vortex line density corresponding to different stages of the tangle's evolution from $L=0$ to the saturated, statistically steady state. The coefficient of the total Andreev reflection was then calculated by averaging Eq.~(\ref{eq:ExAndsimp}) with the values of parameters corresponding to $T=150\,\mu{\rm K}$.

\begin{figure}[t]
    \begin{center}
        \includegraphics[width=0.95\linewidth]{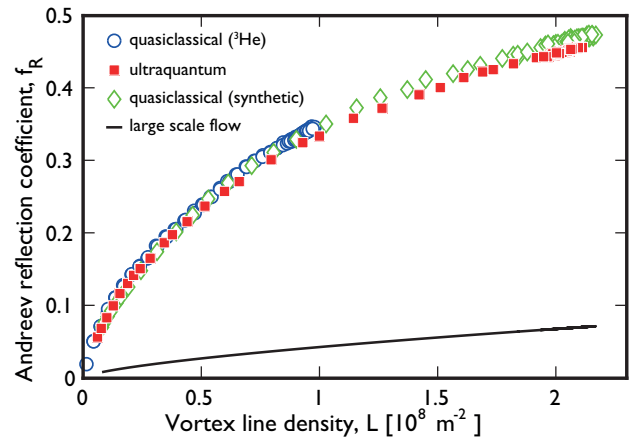}
        \caption{Total Andreev reflection vs vortex line density. Blue circles: $^3$He tangle simulation; red squares: unstructured  (``ultraquantum'') tangle; green diamonds: structured (``quasiclassical'') tangle. Solid black line shows large-scale contribution to the total Andreev reflection, see Eq.~(\ref{eq:AndFracLarge}).}
        \label{fig:CF-KS}
    \end{center}
\end{figure}

The results are shown in Fig.~\ref{fig:CF-KS}. Surprisingly, for the same vortex line density the coefficients of the total Andreev reflection from the unstructured  (``ultraquantum'') and the structured (``quasiclassical'') quantum turbulence appear to be practically indistinguishable, so that the measurements of the total Andreev reflection cannot be used to identify the regime of quantum turbulence~\cite{quasi-vs-ultra}. This result is consistent with experimental observations of Andreev reflection as a function of the grid velocity, see Fig.~\ref{fig:ARvsL} and Refs.~\cite{PhysRevB.95.094518.Jackson,PNAS.111.4659.Fisher,Bradley4} which show that the fully formed vortex tangle can be distinguished from the ``gas'' of individual vortex rings either by the rate of decay of the vortex line density or by the cross-correlation between different detectors.

The results obtained in this and the previous sections show that the total reflection coefficient is mainly determined by the vortex line density, and that the large scale flows emerging in the quasiclassical quantum turbulence have little effect on the total Andreev reflection from the vortex tangle. This conclusion can be further strengthened by estimating the amount of scattering from large scale flows in the quasiclassical regime as follows. We assume that the largest structures generated by the simulations have a size comparable to the box length $D$. For classical turbulence with a Kolmogorov spectrum the characteristic velocity of the flow on a length scale $D$ is $v(D) \approx C_K^{1/2}\epsilon^{1/3}(D/2\pi)^{1/3}$, where $C_K \approx1.6$ is the Kolmogorov constant and $\epsilon$ is the dissipation per unit mass. For quantum turbulence, the dissipation is normally assumed~\cite{VinenNiemela} to be given by $\epsilon = \zeta \kappa^3 L^2$, where the dimensionless constant $\zeta\simeq0.08$ is inferred from experimental data~\cite{Zmeev2015}. The flow generates an energy barrier ${\bf p}_F \cdot {\bf v}(D)$ for incident excitations. Assuming that the barrier is relatively small compared to $k_BT$, the corresponding total reflectivity is $f_l\simeq p_fv(D)/(4k_BT)$ where the factor of 1/4 arises from taking an angular average of the energy barrier. The fraction of quasiparticles reflected by the large scale flow is thus estimated as
\begin{equation}
f_l \simeq \frac{\sqrt{C_K}}{4}\frac{p_F\kappa}{k_BT}\left(\frac{\zeta D}{2 \pi}\right)^{1/3}L^{2/3}\,.
\label{eq:AndFracLarge}
\end{equation}
This is shown by the lower black line in Fig.~\ref{fig:CF-KS}. Even at high line densities the contribution of the large scale flow to the total Andreev reflection is small (only about 15\% at the highest line density, $L\approx2.2\times10^8\,{\rm m}^{-2}$; note, however, that this simple model overestimates the value of $f_l$).

\section{Screening mechanisms in Andreev reflection from vortex tangles}
\label{sec:screening}

In the numerical simulations described in previous sections, as the vortex line density becomes larger than  $\sim2\times10^7\,{\rm m}^{-2}$ the growth of the total Andreev reflection with the vortex line density becomes more gradual, see Fig.~\ref{fig:ARvsL}. The nonlinearity of the growth of the reflection coefficient with $L$ is due to the screening effects which become more important as the density of vortex configuration increases. We use the term `screening' to identify processes which change the overall reflectivity of the tangle for a given line density. Screening, whose mechanisms are discussed below in this section, can occur between different vortices, or between different segments of the same vortex.

Analyzed in detail in Refs.~\cite{BSS1,PNAS.111.4659.Fisher,Sergeev2015}, the Andreev reflection from the flow field of an isolated, rectilinear vortex line is determined by the $1/r$ behavior of the superfluid velocity, where $r$ is the distance from the vortex core~\cite{angle}. For dilute vortex configurations and tangles the total Andreev reflection is a sum of reflections from individual vortices. However, as the line density increases the total reflection no longer remains additive due to screening effects which may either reduce or enhance the Andreev reflectivity of the vortex configuration.

We can identify two main mechanisms of screening. The first, which was called ``partial screening'' in earlier publications~\cite{BSS2,BSS3}, may significantly reduce the total reflectivity of the dense vortex configuration due to a modification of the $1/r$ superflow field in the vicinity of the vortex line. This type of screening may occur between different vortices in a dense vortex configuration, or between different segments of the same vortex. In particular, the reflectivity may be reduced by partial cancellation of the flow produced by neighboring vortices of opposite polarity. For example, the flow field at some distance from two, sufficiently close, (nearly) antiparallel vortex lines is that of a vortex dipole, $v\sim1/r^2$; this, faster decay of the superflow velocity significantly reduces the total reflectivity of the vortex pair.

The $1/r$ velocity field may also be modified by the effect of large curvature of the vortex line itself.  The r\^{o}le of curvature effects has been analyzed in Ref.~\cite{BSS4} for a gas of vortex rings of the same size but random positions and orientations. It was shown that the contribution to partial screening arising from curvature effects may lead to a noticeable reduction of Andreev reflection only for rings whose radii are smaller than $R_*\approx2.4\,\mu{\rm m}$.

It can, therefore, be expected  that partial screening has another, small contribution from effects of large line curvatures arising due to high frequency Kelvin waves propagating along filaments. We may assume that in saturated vortex tangles the additional contribution to partial screening will arise due to Kelvin waves generating line curvatures, $C=\vert d^2{\bf s}/d\xi^2\vert$ (where ${\bf s}$ identifies a point on the vortex line, and $\xi$ is the line's arclength) larger than $C_*=1/R_*\approx4\times10^5\,{\rm m}^{-1}$. From our numerical simulation, described above in Sec.~\ref{sec:vortgeneration}, we calculated the probability distribution function, ${\cal F}(C)$ of line curvatures normalized such that $\int_0^\infty{\cal F}(C)\,dC=1$. Then the fraction of contribution of high-frequency Kelvin waves to Andreev reflection from the tangle can be estimated by
\begin{equation}
\int\limits_{C^*}^\infty{\cal F}(C)\,dC\,.
\label{eq:KWcontrib}
\end{equation}
Numerical estimates of integral~(\ref{eq:KWcontrib}) for the considered saturated tangle of the density $L=9.7\times10^7\,{\rm m}^{-2}$ show that the contribution of high frequency Kelvin waves to partial screening and hence the reduction of the total Andreev reflection from the tangle is negligibly small (less than 1\%)~\cite{KW-contribution}.

The other mechanism is that of `fractional' screening (also called `geometric' screening in Ref.~\cite{PNAS.111.4659.Fisher}). This mechanism is of particular importance for dense or/and large tangles: Vortices located closer to the source of excitations screen those located at the rear. This screening mechanism will also be important in the case where, in dense tangles, vortex bundles may form~\cite{BagLauBar} producing regions of high Andreev reflectivity surrounded by those whose reflectivity is much lower. This will generate a significant `geometric' screening since few excitations can reach vortices at the rear of a large vortex bundle, and adding a vortex to a region where there is already an intense reflection will not have a noticeable effect on the total reflectivity.

Earlier a simple model~\cite{Bradley3} of the fractional screening was proposed to infer a rough estimate for the vortex line density from the experimentally measured Andreev reflection. Assuming that all vortex lines are orthogonal to the beam of excitations, and modeling the Andreev reflection from each vortex as that from an isolated, rectilinear vortex line, the total fraction of excitations, Andreev-reflected by the tangle whose thickness in the direction of quasiparticles' beam is $d$, was found to increase with $d$ as $f_R=1-\exp(-d/\lambda)$, where $\lambda=P^{-1}$ is the lengthscale of exponential decay of the flux transmitted through the tangle and $P=b_0L$ is the probability of reflection per unit distance within the tangle, with $b_0$ given by Eq.~(\ref{eq:AndImpact}) being the scale of Andreev scattering length by a rectilinear vortex. Then the vortex line density can be inferred from the experimentally measured value of $f_R$ as
\begin{equation}
L=-\gamma\ln(1-f_R)\,,
\label{eq:Lvsf}
\end{equation}
where, for the considered model~\cite{Bradley3}, $\gamma=(b_0d)^{-1}$. As noted in Ref.~\cite{Fisher_review}, formula~(\ref{eq:Lvsf}) gives only rough estimates for $L$ which are correct within a factor of 2. This is not surprising since the model~\cite{Bradley3} ignores an influence of geometry and orientation of vortex lines on the Andreev reflection.

\begin{figure}[t]
    \begin{center}
    \includegraphics[width=7.5cm]{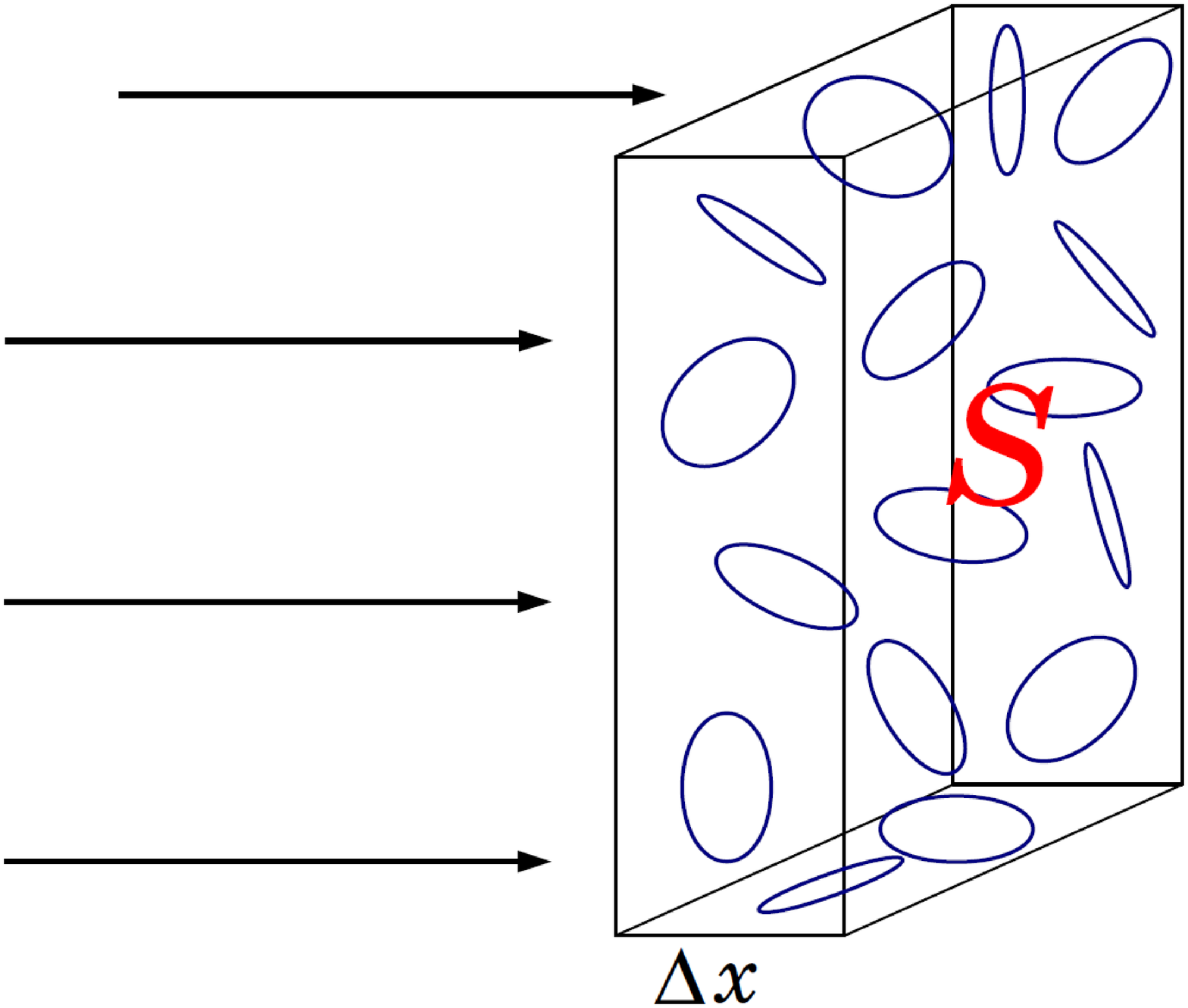}
     \end{center}
    \caption{Modeling the tangle by a configuration of randomly positioned and oriented vortex rings of the same radius. The model is used for estimating the line density from the measurements of Andreev reflection. Black arrows sketch the trajectories of incident excitations.}
    \label{fig:slab}
\end{figure}

This model can be improved by modeling the tangle as a collection of vortex rings of the same radius, $R$, but of random orientations and positions. Andreev reflection from such vortex configurations was analyzed in more detail in our earlier work~\cite{BSS4}. Following Ref.~\cite{Bradley3}, consider a thin slab of superfluid, of thickness $\Delta x$ and cross section $S$, but consisting now of $N$ randomly oriented and positioned vortex rings of radius $R$, see Fig.~\ref{fig:slab}.
The total cross section of Andreev scattering is $\sigma_{tot}=\langle\sigma\rangle N$, where $\langle\sigma\rangle$ is the Andreev cross section of the vortex ring, ensemble-averaged over all possible ring orientations, and $N$ is the total number of vortex rings within the considered slab. In the case where the ring is not too small, $R>R_*\approx2.4\,\mu{\rm m}$, it was found~\cite{BSS4} that $\langle\sigma\rangle\approx2\pi K_lR$, with $K_l=10\,\mu{\rm m}$ at temperature $T=100\,\mu{\rm K}$. The total number of vortex rings within the slab can be calculated as $N=\Lambda/(2\pi R)$, where $\Lambda=LS\Delta x$ is the total line length. The probability that an incident excitation will be Andreev-reflected is $\Delta p=\sigma_{tot}/S=\langle\sigma\rangle L\Delta x/(2\pi R)$, so that the lengthscale of decay the quasiparticle flux is $\lambda=\Delta x/\Delta p=1/(K_lL)$. Then the line density inferred from the fraction of excitations Andreev-reflected by the tangle of thickness $d$ is given by formula~(\ref{eq:Lvsf}), where now
\begin{equation}
\gamma=(K_ld)^{-1}\,.
\label{eq:gamma}
\end{equation}
For the considered temperature, we find $b_0\approx6.3\,\mu{\rm m}$ and $K_l\approx10\,\mu{\rm m}$, so that formula~(\ref{eq:Lvsf}) with $\gamma$ defined by Eq.~(\ref{eq:gamma}) provides a more accurate estimate yielding values of $L$ which, for the same value of Andreev reflection $f_R$, are about twice smaller than those predicted by the original model of Bradley \textit{et al.}~\cite{Bradley3} (however, note our comment in Ref.~\cite{L-Kl}).

\section{Spectral properties of Andreev reflection vs those of quantum turbulence}
\label{sec:fluctuations}

\begin{figure}[t]
    \begin{center}
        \includegraphics[width=0.95\linewidth]{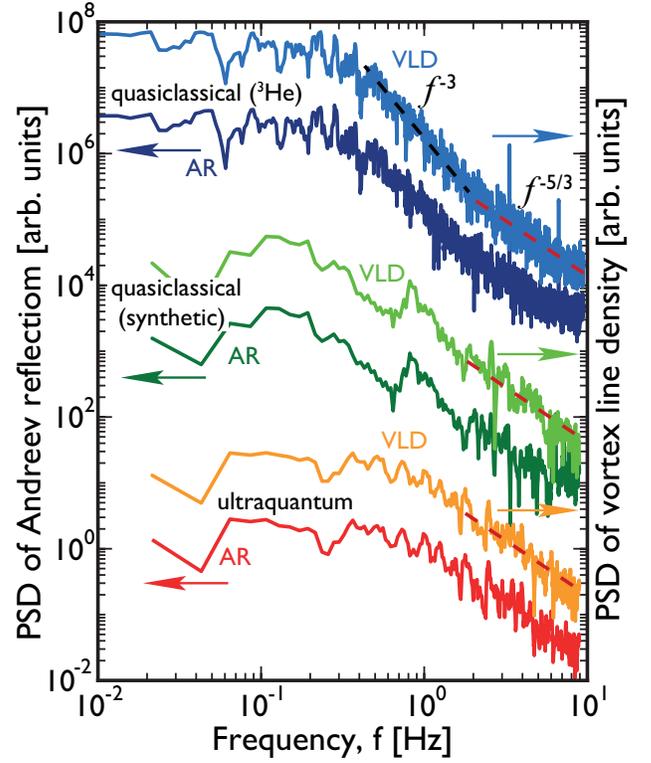}
        \caption{Power spectral densities of simulated Andreev reflection and simulated vortex line density corresponding to (from the top to bottom): quasiclassical turbulence in $^3$He (light blue: vortex line density, dark blue: Andreev reflection); synthetic turbulence (light green: vortex line density, dark green: Andreev reflection); counterflow turbulence (orange: vortex line density, red: Andreev reflection). The red and black dashed lines are guides to the eye for the $f^{-5/3}$ and $f^{-3}$ scaling respectively.}
        \label{fig:anddensfluc}
    \end{center}
\end{figure}

In the simulation of quasiclassical turbulence using ring injection (see Fig.~\ref{fig:transient} in Sec.~\ref{sec:vortgeneration}), once the tangle has reached the statistically steady state, the vortex line density and the Andreev reflection coefficient fluctuate around their equilibrium, time-averaged values, $\langle L\rangle=9.7\times10^{7}\,{\rm m}^{-2}$ and $\langle f_R\rangle=0.34$, respectively. These fluctuations contain important information about the quantum turbulence. In order to compare the spectral characteristics of fluctuations of the Andreev reflection and the vortex line density,
\begin{equation}
\delta f_R(t)=f_R(t)-\langle f_R\rangle\,, \quad \delta L(t)=L(t)-\langle L\rangle\,,
\label{eq:delta-RL}
\end{equation}
respectively, we monitor a steady state of simulated tangle for a period of approximately 380\,s or 7500 snapshots. Taking the Fourier transform $\widehat{\delta f_R}(f)$ of the time signal $\delta f_R(t)$, where $f$ is frequency, we compute the power spectral density $\vert \widehat{\delta f_R}(f) \vert^2$ (PSD) of the Andreev reflection fluctuations. Similarly we compute the PSD $\vert \widehat{\delta L}(f) \vert^2$ of the vortex line density fluctuations and plot both of them on Fig.~\ref{fig:anddensfluc}.

Figure~\ref{fig:anddensfluc} also shows the PSDs of Andreev reflection fluctuations and of vortex line density fluctuations for mechanically driven (synthetic quasiclassical) and counterflow (ultraquantum) tangles. For each of the simulations, the PSDs of Andreev reflection and vortex line density fluctuations look very similar. A difference between the PSDs for our $^3$He tangle simulated using vortex ring injection is somewhat larger and may be caused by the presence of large scale flows that affect the spectral properties of the fluctuations, but not the time-averaged values of the Andreev reflection and vortex line densities.

\begin{figure}[t]
    \begin{center}
        \includegraphics[width=0.95\linewidth]{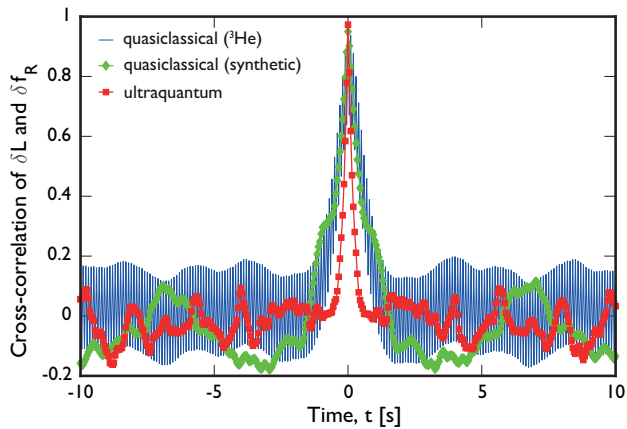}
        \caption{Cross-correlation between the fluctuations of Andreev reflection and vortex line density for various simulated tangles. The value of the cross-correlation peak exceeds 0.9 and shows strong link between the fluctuations of Andreev reflection and vortex line density.}
        \label{fig:cross-correlation}
    \end{center}
\end{figure}

Our observations on the similarities of the PSD of Andreev reflection and vortex line density fluctuations are supported by the normalized cross-correlations between the fluctuations of the vortex line density, $\delta L(t)$, and the fluctuations of the Andreev reflection, $\delta f_R(t)$, which is computed as
\begin{equation}
    F_{LR}(\tau)=\frac{\langle \delta L(t)\delta f_R(t+\tau) \rangle}
    {\sqrt{\langle\delta L^2(0) \rangle}\sqrt{\langle\delta f_R^2(0) \rangle}}\,,
    \label{eq:correl}
\end{equation}
where the angle brackets indicate averaging over time, $t$, in the saturated regime, and $\tau$ is the time-lag. Figure~\ref{fig:cross-correlation} shows the cross-correlation between the vortex line density and the Andreev reflection for the simulated tangles. The cross-correlation values $F_{LR}(0)$ for ultraquantum and synthetically generated tangles approach unity, while for $^3$He tangle is slightly lower, $F_{LR}(0)\approx0.9$. The high values of the cross-correlation clearly demonstrate a strong link between the fluctuations of Andreev reflection and vortex line density.

We will now analyze the frequency dependence of the PSDs of the vortex line density fluctuations shown in Fig.~\ref{fig:anddensfluc}. At high frequencies, the vortex line density fluctuation spectrum for all of the simulated tangles exhibits $f^{-5/3}$ scaling and is dominated by the contribution of unpolarized, random vortex lines~\cite{BagLauBar}. In the intermediate frequency range, only the quasiclassical $^3$He fluctuation spectrum shows a clear $f^{-3}$ scaling, which is governed by the large scale flows indicating polarized (that is, cooperatively correlated) vortex lines in agreement with the recent numerical simulations~\cite{BagLauBar}. If we reasonably assume that a crossover between $f^{-3}$ to the $f^{-5/3}$ behavior should occur at around the frequency corresponding to the intervortex distance, $\ell$, then using the value of 102$\,\mu$m for $\ell$ calculated above for the equilibrium tangle, the crossover should occur at frequency $f_{\ell} \approx v/\ell=\kappa/(2 \pi \ell)\approx1$\,Hz. As seen in Fig.~\ref{fig:anddensfluc}, this is in a very fair agreement with the frequency $\approx2$\,Hz of the crossover between the two regimes.

\begin{figure}[ht]
    \begin{center}
        \includegraphics[width=0.95\linewidth]{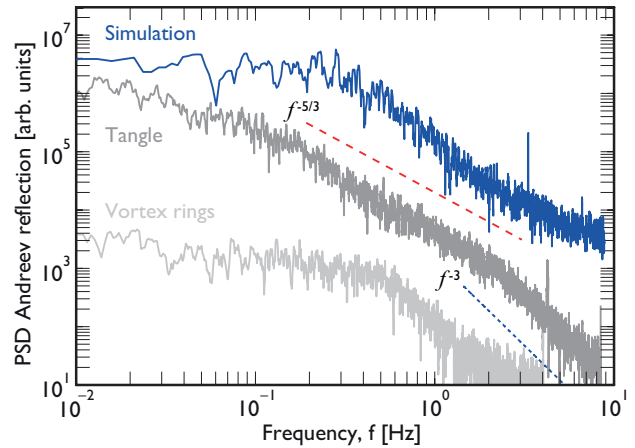}
        \caption{Power spectral density of the Andreev reflection for numerical simulations (top, blue) and experimental observations\cite{PRL.115.015302.Baggaley} (middle and bottom, gray) versus frequency. The red and blue dashed lines are guides to the eye for the $f^{-5/3}$ and $f^{-3}$ scaling respectively. For the details, see text. }
        \label{fig:flucsimtanglerings}
    \end{center}
\end{figure}

Figure~\ref{fig:flucsimtanglerings} shows and compares the PSDs of fluctuations of Andreev reflection for the simulation (top, blue) and for the experiments\cite{PRL.115.015302.Baggaley} (middle and bottom, gray). The experimental data are shown for a fully-developed tangle (dark gray) and ballistic vortex rings (light gray) corresponding to the grid velocities of 6.3\,mm\,s$^{-1}$ and 1.9\,mm\,s$^{-1}$ respectively. (The prominent peaks in the numerical data are artifacts of the discrete vortex-ring injection process.) The power spectrum of $\delta f_R(t)$ of the simulation and of the experimental data for the developed tangle (reported here and in Refs.~\cite{Bradley4,PRL.115.015302.Baggaley}) are in overall agreement showing reduction of the signal with increasing frequency.

A detailed comparison of experimental observations at high grid velocity and simulations can be based on two alternative approaches to the interpretation of numerical and experimental data. Both of these approaches are viable, mostly due to a rather noisy character of the fluctuating Andreev reflection's spectra for simulated tangles, but they lead to somewhat different conclusions. In the first approach\cite{PRL.115.015302.Baggaley} we could identify the same $f^{-5/3}$ scaling behavior at intermediate frequencies. This scaling, observed experimentally in high temperature $^4$He \cite{Roche2007} and reproduced by numerical simulations \cite{BagLauBar}, has been interpreted \cite{Roche2008} as that corresponding to the fluctuations of passive (material) lines in Kolmogorov turbulence. At high frequencies the experimental data develop a much steeper scaling ($\approx\,f^{-3}$), not seen in the numerical spectrum, probably owing to the finite numerical resolution. However, this frequency dependence {\it is} observed in the experiment in the case where only microscopic vortex rings are propagating through the active region and there are no large-scale flows or structures. Thus, we can argue that the $f^{-3}$ scaling for the vortex tangles corresponds to Andreev reflection from superflows on length scales smaller than the intervortex distance.

\begin{figure}[ht]
    \begin{center}
        \includegraphics[width=0.95\linewidth]{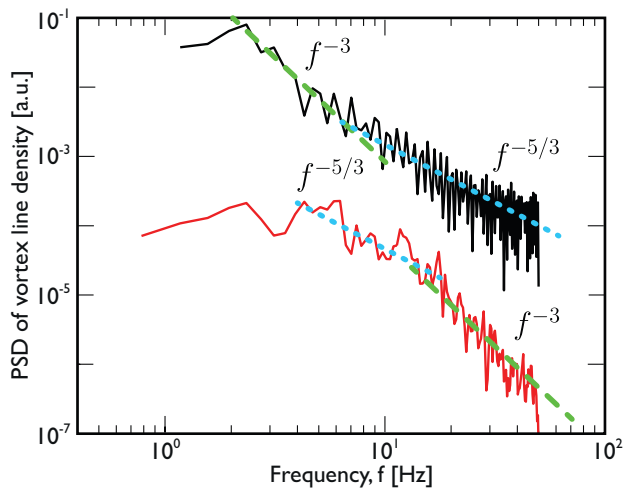}
        \caption{Power spectral density of fluctuations of the total vortex line density (black upper solid line) and of the polarized vortex line density (red lower solid line)~\cite{BagLauBar}. The light blue dotted and green dashed lines are guides to the eye for the $f^{-5/3}$ and $f^{-3}$ scaling respectively. For the details, see text. }
        \label{fig:polarised}
    \end{center}
\end{figure}

In the second, alternative approach we assume that the PSD of Andreev reflection fluctuations mimics the functional dependence of the PSD of the vortex line density in accordance with Fig.~\ref{fig:flucsimtanglerings}. In this case the Andreev-reflection fluctuation spectra exhibit a $f^{-5/3}$ dependence at high frequency and a steeper ($\approx\,f^{-3}$) dependence at intermediate frequencies, which is almost opposite to the scaling observed in the experiment. Figure~\ref{fig:polarised} shows fluctuation spectra of the vortex line density~\cite{BagLauBar} calculated for the total line density (black upper solid line) and the polarized fraction  of the quasiclassical tangle (red lower solid line). The observed frequency dependencies are similar, and suggest that in our experiments the vortices probed by quasiparticles are polarized and hence the tangle generated by the vibrating grid is much more polarized than expected. It is feasible that vortex rings moving predominantly in the same direction polarize the resulting tangle, and that quasiparticles accompanying turbulence production further assist the polarization of the tangle in the direction of grid motion.

\section{Discussion and conclusions}
\label{sec:interpretations}

In this study we combined the three-dimensional numerical simulations of dynamics and evolution of turbulent vortex tangles, numerical analysis of Andreev scattering of thermal quasiparticle excitations by the simulated tangles, and experimental measurements of Andreev reflection by quantum turbulence in $^3$He-B at ultralow temperatures. Our aim was to show that a combined study is a powerful tool that might reveal important properties of pure quantum turbulence.

We have shown that, with sufficient resolution of the Andreev retroreflected signal, the Andreev scattering technique can generate a map of local reflections which reveals a structure of the vortex tangle. Furthermore, in the numerical experiment it is possible to separate the quasiparticle and quasihole excitations. The individual maps of Andreev reflection for each of these species of excitations also show the presence and the structure of large scale flows arising in the quasiclassical (Kolmogorov) regime of quantum turbulence. However, in the real physical experiment the thermal beam consists of excitations of both sorts whose separation is impossible. We found that, while a presence of large scale flows may still be detected by the enhancement of local Andreev reflections, an identification of the detailed large scale flow structure would hardly be possible.

For the quasiclassical vortex tangle, we calculated the total Andreev reflection as a function of the vortex line density. We found that the results of our numerical simulation reproduce rather well the experimental measurements of the total Andreev reflection, not only qualitatively but also quantitatively, although this good quantitative agreement should be taken with some caution for the reason explained in Sec.~\ref{sec:exptl}. We also found that the total reflection is strongly affected by the screening effects. Here we use the term ``screening'' to identify mechanisms which, for a given line density, change the overall reflectivity of the tangle. In particular, screening mechanisms are responsible for the slowing of growth of the total reflectivity with the vortex line density when the latter reaches, in our simulation, the value about $2\times10^7\,{\rm m}^{-2}$. We identified two main mechanisms of screening. The first, so-called ``partial screening'' may significantly reduce the total reflectivity of the dense vortex configuration due to a modification of the $1/r$ superflow in the vicinity of the vortex line. The second is the `fractional' (or `geometric') screening and is of importance for dense or/and large tangles: Vortices located closer to the source of excitations screen those at the rear. We investigated screening mechanisms in some detail and conclude that they often should be taken into consideration in interpretation of experimental observations and numerical simulations.

While this paper is mostly concerned with the Andreev reflection from quasiclassical vortex tangles, there also exists another regime of quantum turbulence of a very different spectral nature in which the energy is contained in the intermediate range of scales and the energy spectrum nowhere conforms to the Kolmogorov $-5/3$ scaling. Unlike quasiclassical quantum turbulence, this form of turbulence, named `ultraquantum' or `Vinen' turbulence~\cite{Volovik}, is unstructured: the only length scale in the ultraquantum tangle is the mean intervortex separation. At very low temperatures the ultraquantum regime of quantum turbulence has been identified experimentally in $^4$He~\cite{Walmsley-2008,Zmeev2015} and $^3$He-B~\cite{Bradley3}. Considering that the Andreev reflection can reveal the presence of the large scale flows existing in quasiclassical turbulence, it could be expected that the ultraquantum tangle can be distinguished from the quasiclassical one by their total Andreev reflectivities. However, our modeling study in Sec.~\ref{sec:ultraquantum} of Andreev reflection from the structured (Kolmogorov) and unstructured (ultraquantum) vortex tangles for the same vortex line density has demonstrated that this is not the case, and that the total reflection is determined by the vortex line density alone. This, in particular, means that the contribution of the large scale flow to the total Andreev reflection is small.

Finally, we investigated the spectral characteristics of Andreev reflection from various tangles and compared them with the spectral characteristics of the quantum turbulence. We found that the fluctuations of the vortex line density and of the Andreev reflection are very strongly correlated and also have similar fluctuation power spectra. The Andreev-reflection fluctuation spectra and of the vortex line density fluctuations look practically identical in the case of ultraquantum turbulence but may differ for the quasiclassical tangle, that is when a large scale flow is present. Therefore, some caution should be taken in interpreting the $f^{-5/3}$ scaling of the Andreev retroreflected signal's frequency spectrum, observed in the experiment of Bradley {\it et al.}~\cite{Bradley4}, as a direct representation of the spectral properties of the vortex-line density fluctuations in turbulent $^3$He-B. Our results show that this scaling could be attributed to the fluctuations of passive (material) lines in Kolmogorov turbulence, or to the fluctuations of the vortex line density of a strongly polarized tangle. The latter possibility suggests that a turbulent tangle produced by the vibrating grid at high velocities is more polarized than has usually been assumed earlier. It is very clear that the Andreev reflection technique has great potential for elucidating the properties of pure quantum turbulence, but requires further work. 

It is clear that a useful study could be made of a tangle created from rings emitted by two grids facing each other. Such a tangle should have a vortex line density similar to that of the fully-developed tangle studied in our work but with a much smaller polarization. Numerically, it would be interesting to investigate fluctuations of the Andreev reflection and vortex line density for the shower of ballistic vortex rings using a higher spatial resolution and whether the $f^{-3}$ spectrum would reproduced experimentally. The influence of large scale flows on Andreev reflections should also be studied in superfluid $^3$He under rotation since this scenario uniquely offers a vortex line ensemble with inherent strong polarization.

\begin{acknowledgments}
This work was supported by the Leverhulme Trust grants No. F/00 125/AH and F/00 125/AD, the European FP7 Program MICROKELVIN Project 228464, and the UK EPSRC grants No. EP/L000016/1, No. EP/I028285/1 and EP/I019413/1. MJJ acknowledges the support from Czech Science Foundation Project No. GA\v{C}R 17-03572S.

All data used in this paper are available at http://dx.doi.org/10.17635/lancaster/researchdata/xxx, including descriptions of the data sets.
\end{acknowledgments}


\begin{thebibliography}{99}
\bibitem{timescale}
Because the time scale of quasiparticle motion is much smaller than both the time scale of the evolution of the tangle and the time scale of the fluid motion within the tangle, $\tau_f\sim1/\sqrt{\kappa L}$, calculation of the Andreev reflection at any time during the transient were performed for the tangle frozen at this moment of time.
\bibitem{fvsL}
Changing the injection and switching frequencies, as well as the injected ring's size, we also generated the tangles whose time-averaged, saturated line densities were the same as some of the line densities during the transient. Calculations of the fraction of Andreev-reflected excitations for such tangles yielded the results consistent with those shown in the top panel of Fig.~\ref{fig:ARvsL}.
\bibitem{smallL}
The injected rings are relatively large so that the effects of high curvatures on the Andreev reflection can be neglected~\cite{BSS4}.
\bibitem{KW-damping}
Some dissimilarity between these tangles and the tangles simulated for the truly zero temperature limit is that at elevated temperatures the Kelvin waves are damped by the mutual friction. However, as will be noted in Sec.~\ref{sec:screening} Kelvin waves have a negligible effect on the total Andreev reflection.
\bibitem{quasi-vs-ultra}
The question whether the regime of quantum turbulence, quasiclassical or ultraquantum, can be identified by spectral properties of Andreev reflection, still remains open. For quasiclassical regime a correspondence between the spectral properties of Andreev reflection and those of quantum turbulence are discussed in Ref.~\cite{PRL.115.015302.Baggaley} and in Sec.~\ref{sec:fluctuations} of this paper.
\bibitem{angle}
In the cited papers the analysis of Andreev reflection was given for the case where the beam of incident excitations is orthogonal to the vortex line. In the general case the dependence of Andreev reflection on the angle $\theta$ between the direction of the incident beam and the vortex line should be accounted for by a factor $\sin\theta$ in the formulae for a fraction of Andreev reflected excitations and the corresponding scattering length.
\bibitem{KW-contribution}
It should be noted that although Kelvin waves of wavelength smaller that $R_*$ do not contribute significantly to the Andreev reflection, they may add to the total line density of the tangle.
\bibitem{L-Kl}
The values of the vortex line density found from formula~(\ref{eq:Lvsf})
with $\gamma$ defined by Eq.~(\ref{eq:gamma}) may be underestimated
for very dense tangles or configurations of rings such that overlapping
of rings' flow fields (and hence partial screening) can no longer be
neglected. For a more detailed analysis of Andreev reflection
from dense configurations of interacting vortex rings see our
earlier paper~\cite{BSS4}.
\end{thebibliography}
\end{document}